\documentclass[sigconf]{acmart}

\usepackage{booktabs} 
\usepackage{hyperref}
\usepackage{balance} 

\setcopyright{rightsretained}

\acmDOI{}

\acmISBN{}

\acmConference[PCSC2025]{Philippine Computing Science Congress}{May 2025}{Benguet, Philippines}
\acmYear{2025}
\copyrightyear{2025}

\acmArticle{}
\acmPrice{}

\acmBooktitle{Proceedings of the Philippine Computing Science Congress (PCSC 2025), Benguet, Philippines}
\editor{}
\editor{}
\editor{}

\begin{document}

\title[SpeakAR]{Speak with Confidence: Designing an Augmented Reality Training Tool for Public Speaking}

\author{Mark Edison {Jim}}
\authornote{Both authors contributed equally to this research.}
\affiliation{%
  \institution{De La Salle University}
 \city{Manila}
  \country{Philippines}}
\email{mark_edison_jim@dlsu.edu.ph}

\author{Jan Benjamin {Yap}}
\authornotemark[1]  
\affiliation{%
  \institution{De La Salle University}
 \city{Manila}
  \country{Philippines}}
\email{jan_benjamin_yap@dlsu.edu.ph}

\author{Gian Chill {Laolao}}
\affiliation{%
  \institution{De La Salle University}
 \city{Manila}
  \country{Philippines}}
\email{gian_laolao@dlsu.edu.ph}

\author{Andrei Zachary {Lim}}
\affiliation{%
  \institution{De La Salle University}
 \city{Manila}
  \country{Philippines}}
\email{andrei_zachary_lim@dlsu.edu.ph}

\author{Jordan Aiko {Deja}}
\orcid{0001-9341-6088}
 \affiliation{%
  \institution{De La Salle University}
  \city{Manila}
  \country{Philippines}}
\email{jordan.deja@dlsu.edu.ph}

\renewcommand{\shortauthors}{Jim and Yap et al.}

\begin{abstract}
Public speaking anxiety affects many individuals, yet opportunities for real-world practice remain limited. This study explores how augmented reality (AR) can provide an accessible training environment for public speaking. Drawing from literature on public speaking, VR-based training, self-efficacy, and behavioral feedback mechanisms, we designed SpeakAR, an AR-based tool that simulates audience interaction through virtual models. SpeakAR was evaluated with five participants of varying anxiety levels, each completing six speaking tasks. Results indicate that AR exposure can enhance confidence, with participants finding the system useful for practice. Feedback highlighted the importance of dynamic facial expressions and idle animations in virtual models to improve realism and engagement. Our findings contribute to the design of AR-based training tools for public speaking, offering insights into how immersive environments can support skill development and anxiety reduction.

\end{abstract}

%
%


\keywords{augmented reality, speaker, avatar, mobile, design, storyboarding, \texttt{Unity}, public speaking}

\begin{teaserfigure}
  \includegraphics[width=\textwidth]{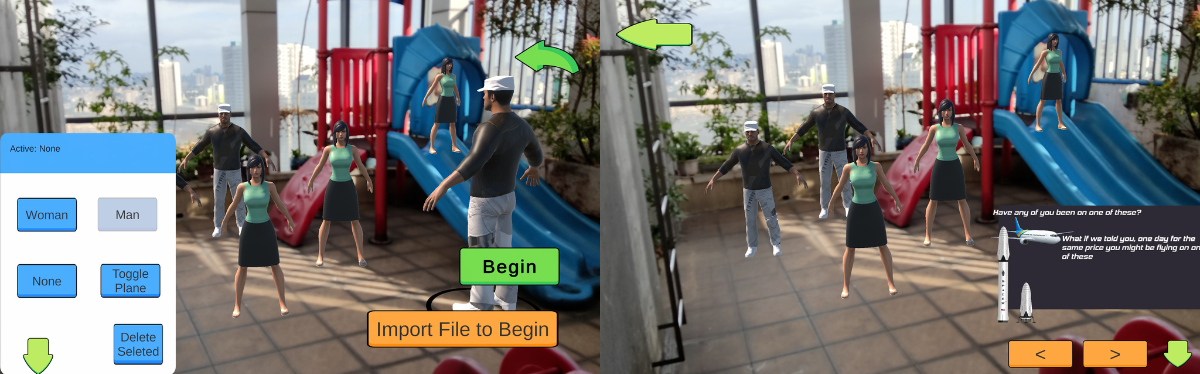}
  \caption{We introduce SpeakAR, an augmented reality training tool for public speaking. Designed for mobile devices, it renders digital avatars in a real-world environment to support users facing speaking challenges, such as glossophobia. Users can customize their audience setup (as shown on the left), adjusting avatar placement and appearance. Additionally, SpeakAR enhances immersion by seamlessly integrating presentation slides into the user’s view, allowing them to rehearse with a virtual audience without breaking focus.}
  \label{fig:teaser}
\end{teaserfigure}

\maketitle

\section{Introduction}
\par Public speaking is not a skill that is given or a default to everyone. Some people usually struggle with speaking in front of an audience and will take some time to get used to it. Glossophobia, the fear of public speaking, affects approximately 77\% of the population \cite{Heeren:2013:APS, dansieh2021glossophobia, dellah2020glossophobia}. A study conducted at St. Paul University Surigao by Plandano et al. \cite{Plandano:2023:PSA} found that 78\% of surveyed students experience this fear, often citing a lack of confidence as the primary cause. Over the years, various methods have been explored to alleviate public speaking anxiety, including computer-based training (CBT). CBT leverages technology to provide accessible, self-paced learning environments that support skill development without the constraints of traditional in-person training.

\par Among the digital approaches, Extended Reality (XR) technologies — including Virtual Reality (VR) and Augmented Reality (AR) — have emerged as promising tools for public speaking training. VR enables users to practice in fully immersive, simulated environments that replicate real-world speaking scenarios, providing a controlled space for exposure therapy \cite{Anderson:2005:CBT}. AR, on the other hand, overlays digital elements onto the physical world, allowing users to engage with virtual audiences while maintaining a connection to their real-world surroundings. Compared to traditional training, XR-based simulations offer key advantages such as controllability, accessibility, and adaptability \cite{Lo:2022:EEF, yoneda2024xr}.

\par While numerous VR-based applications for public speaking exist, their implementations vary significantly in usability and effectiveness. Additionally, AR remains underexplored as a medium for public speaking training despite its potential benefits. This paper introduces \textbf{SpeakAR}, an AR-based tool designed to help users practice public speaking in an accessible and flexible environment. By simulating audience interaction through virtual models, SpeakAR provides real-time feedback to enhance confidence and reduce anxiety. We evaluate SpeakAR’s efficacy through a user study, analyzing how AR-based interventions can impact public speaking skills. Our findings may contribute to the broader understanding of XR-based training tools and their role in addressing conditions such as glossophobia. In summary, we summarize our contributions as follows: 
\begin{itemize}
    \item \textbf{Design}. We inform the design of an augmented reality (AR) training tool for public speaking drawing on the current state-of-the-art and building on lessons learned from these studies. 
    \item \textbf{Prototype}. We developed and implemented an AR training tool which we refer to as \textbf{SpeakAR}. Its features were inspired by the design elements we presented in our previous contribution. 
    \item \textbf{Empirical}. We report the findings of a small pilot study that evaluates its initial effects. While still at its early stages, findings can be used towards the design of a more comprehensive training tool and a more controlled protocol to evaluate its long-term effects. 
\end{itemize}

\section{Background and Related Work}
\subsection{Related Systems}
\par Several VR-based public speaking training systems have been developed with varying features such as the works of ~\cite{bachmann2023virtual, zhou2021virtual}. Palmas et al. \cite{Palmas:2019:AVR} created an application using the Unity3D engine, incorporating speech content analysis, detection of filler words, and body language tracking. Their system included semi-realistic virtual audiences that adjusted their mood based on the speaker's words per minute, volume, eye contact, and posture. An enhancement introduced in their 2021 study \cite{Palmas:2021:VRPST} featured a direct feedback system that displayed real-time performance indicators. These indicators, such as colored symbols reflecting audience engagement, helped users adjust their speaking style dynamically.

\par Frisby et al. \cite{Frisby:2020:UVR}, in contrast, used a simpler approach, presenting students with a looping 360-degree video of a classroom environment. This method provided an immersive but passive experience. Truong et al. \cite{Truong:2022:PPS} utilized the Windows Speech Library for speech transcription and analysis, identifying unique words, stop words, and stuttering patterns. Additionally, they incorporated audience feedback by tracking users' gaze and mapping facial expressions to emotional scores.

\par El Yamri et al. \cite{El:2019:ERA} developed a system that adjusted crowd animations based on the speaker’s voice projection. Their application also allowed users to practice public speaking in various environments, such as auditoriums and job interview settings, extending beyond traditional classroom-based simulations. \textbf{Our work builds on these techniques by implementing a tablet-first, mobile-friendly augmented reality training tool that allows users to place avatars within their field of view during practice sessions.} 

\subsection{Effects of AR on Public Speaking}
\par Virtual Reality (VR) has been widely used to explore its impact on public speaking skills. Several studies have examined different aspects of VR-based public speaking training. Several works \cite{Palmas:2019:AVR, powers2008virtual, krijn2004virtual, price2007role} investigated whether Virtual Reality Exposure Therapy could serve as a safe and effective learning environment by simulating realistic public speaking scenarios. In a follow-up study, Palmas et al. \cite{Palmas:2021:VRPST} enhanced their framework by incorporating gamification elements to improve engagement. Similarly, Frisby et al. \cite{Frisby:2020:UVR} evaluated the effectiveness of VR in helping students pass the Basic Communication Course at the University of Kentucky, emphasizing public speaking as a key competency. 

\par In addition to public speaking training, researchers have explored how VR can enhance voice analysis systems. Truong et al. \cite{Truong:2022:PPS} developed a system that integrates speech analysis and audience feedback to assess speech patterns, including stuttering frequencies. El Yamri et al. \cite{El:2019:ERA} focused on improving a pre-existing voice analysis system by simulating crowd reactions based on voice input, although their study did not evaluate direct improvements in public speaking skills. \textbf{Our work builds on the success of existing AR speaking tools by leveraging virtual 3D avatars to help learners maintain focus and engagement. This approach allows users to acclimate to speaking in front of an audience while reducing anxiety associated with real-time crowd reactions.}

\subsection{Effects on User Experience}
\par User experience considerations played a critical role in the design of these systems. Palmas et al. \cite{Palmas:2019:AVR} allowed users to select avatars that matched their gender, improving self-identification and immersion. Their 2021 system \cite{Palmas:2021:VRPST} further optimized user experience by incorporating an intuitive direct feedback mechanism, where performance indicators provided real-time insights without disrupting speech delivery. Users appreciated this gamification approach, reporting increased motivation to improve their speaking performance.

\par Frisby et al. \cite{Frisby:2020:UVR} provided immediate instructor feedback, which participants valued, but noted limitations in accessing presentation aids within the VR environment. Truong et al. \cite{Truong:2022:PPS} tested three versions of their system—Basic (virtual environment only), Lite (with speech analysis), and Full (including audience feedback). Participants preferred the Full version, highlighting its ability to enhance awareness of their speaking environment. Users suggested adding body language tracking and session recording for further improvements. 

\par While most studies explored user experience and engagement, El Yamri et al. \cite{El:2019:ERA} did not specifically evaluate user feedback, focusing instead on the technical improvements of their voice analysis system. 

\par Overall, user responses across studies indicated positive reception towards VR-based public speaking training. Frisby et al. \cite{Frisby:2020:UVR} found that students valued real-time feedback from instructors. Palmas et al. \cite{Palmas:2019:AVR, Palmas:2021:VRPST} demonstrated that prolonged VR exposure could lead to improved public speaking confidence, with users favoring the 2021 gamified version. Truong et al. \cite{Truong:2022:PPS} received favorable reviews, with suggestions for increasing realism through advanced facial expressions and personalized audience avatars. \textbf{To effectively evaluate SpeakAR, we will adopt similar variables used in previous studies, focusing on how user experience influences familiarity and improvement during practice sessions with virtual avatars. However, we will not yet assess overall effectiveness, as this would require more structured and well-controlled evaluation studies.}

\subsection{Summary}
\par While existing VR-based solutions have demonstrated promise in providing immersive practice environments, they come with limitations. The setup process can be time-consuming, and fully capturing the nuances of speech, gestures, and audience reactions remains a challenge. Current speech analysis tools primarily focus on quantifiable features such as speech volume, pace, and hesitation patterns, but they lack the ability to holistically assess speech content and rhetorical effectiveness. Similarly, body language tracking is often limited to predefined gestures, without the ability to fully interpret dynamic speaker-audience interactions. \textbf{This is why we have chosen a tablet-first mobile augmented reality solution over an immersive VR approach, as our goal is to maintain the user’s natural field of view during rehearsals.}

\section{SpeakAR}
\begin{figure}[ht]
    \centering
    \includegraphics[width=0.87\columnwidth]{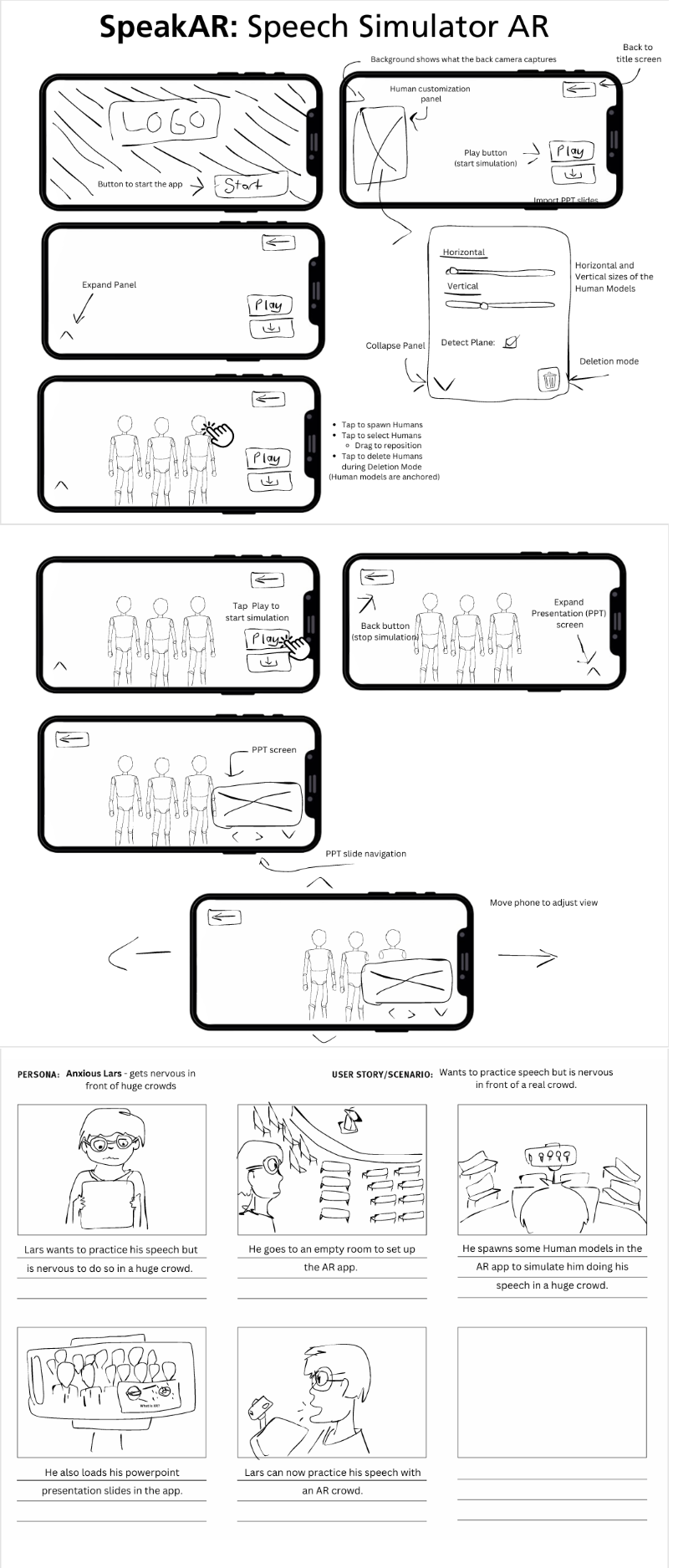}
    \caption{From Sketches to Storyboard. SpeakAR was iteratively designed using templates from XR prototyping \href{https://medium.com/cinematicvr/a-storyboard-for-virtual-reality-fa000a9b4497}{(available here)}. The initial sketches enabled designers to incorporate key features inspired by prior research, guiding the development of assets and environments. These sketches were then translated into a structured storyboard (bottom), which helped refine the screen flow and overall user experience.}
    \label{fig:sketches}
\end{figure}

\subsection{Concept and Rationale}
\par The design of SpeakAR is informed by key insights drawn from prior research on virtual environments for public speaking training. These studies emphasize three critical aspects of an effective training system: (1) real-time analysis of speech content and delivery, (2) tracking of gestures and body language, and (3) immediate and contextually relevant audience feedback. However, we have designed SpeakAR to focus on the ability to replicate realistic speaking environments is crucial to ensuring that users are immersed in a setting that closely resembles real-world public speaking scenarios. SpeakAR is designed to address these gaps by leveraging Augmented Reality (AR) to create a more flexible and accessible public speaking training tool. Unlike VR, which fully immerses users in a digital space, AR overlays interactive feedback mechanisms onto a user’s real-world environment, allowing for a seamless blend of virtual audience response with the user's natural surroundings. This approach minimizes setup complexity and enables users to practice in the actual locations where they will present. The ultimate goal of SpeakAR is to reduce public speaking anxiety by providing an accessible, scalable, and data-driven training solution. 

\subsection{Design Features}
\par We conceptualized SpeakAR, an AR-based public speaking training tool, drawing inspiration from prior works. Since most XR public speaking applications rely on VR, we explored an AR alternative to leverage its portability and accessibility, particularly for users who may not have access to VR headsets. To inform our design, we followed established XR prototyping methods, starting with sketches and storyboards to visualize key use cases and interactions (see ~\autoref{fig:sketches}).

\par We first created a storyboard illustrating how users would engage with SpeakAR in real-world scenarios. This helped us refine the application's core features and user flow. Next, we produced interface sketches to outline the layout, interactions, and feedback mechanisms. These early design artifacts guided our iterative process in defining the application's functionality.

\par For the visual design, we focused on branding, backgrounds, and color schemes. We designed a minimalistic logo that integrated the app name, SpeakAR, a play on the word “speaker” while emphasizing its AR modality. The primary background element—a curtain on the title screen—was chosen to evoke the experience of stepping onto a stage. To enhance usability, we selected a clean white background for the menu, ensuring clarity, while vibrant colors were used for buttons to draw attention without overwhelming the user.

\subsection{Main Features}
\begin{figure*}[ht]
    \centering
    \includegraphics[width=1\textwidth]{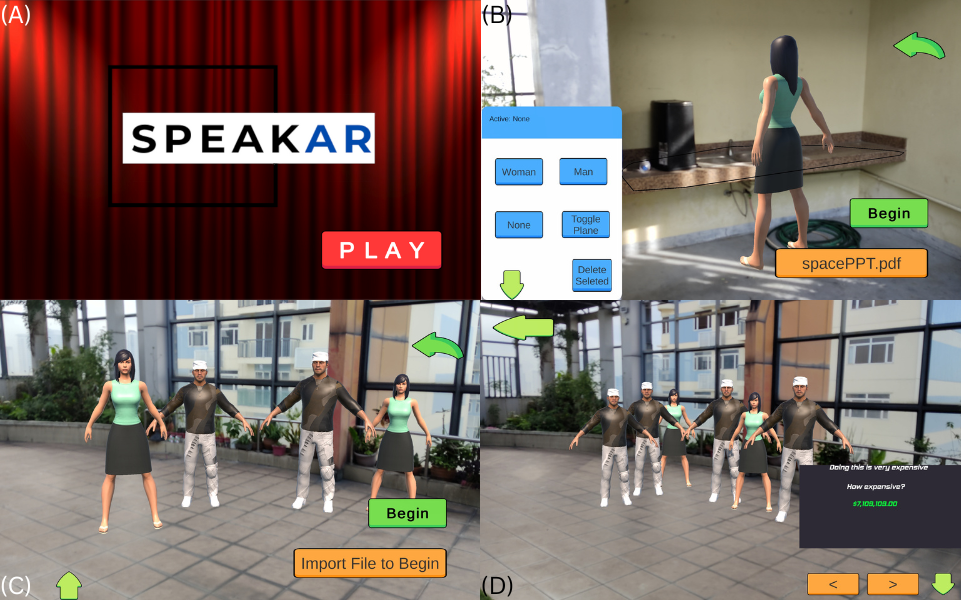}
    \caption{Screenshots showcasing the core features of SpeakAR, a tablet-first application that also functions on mobile phones. (A) The splash screen features curtains that gradually unveil as the user presses Start, reinforcing the stage performance metaphor. The background uses the live camera feed, allowing users to see their own point of view. (B) Users can select and customize their audience avatars, adjusting their type, rotation, and visibility. They can also clear or delete previously placed avatars. (C) Avatars are positioned within the camera view, rendered with realistic lighting to enhance immersion. (C-D) Once the audience setup is complete, users can import a PowerPoint file directly into the AR environment. This seamless integration allows them to practice public speaking while maintaining immersion.}
     \label{fig:allfeatures}
\end{figure*}

\par We developed SpeakAR (see ~\autoref{fig:allfeatures}) in Unity for Android devices, such as smartphones and tablets, to ensure portability and ease of use. The system includes two main features designed to help users practice for speaking engagements: audience placement and visual aid access.

\par The audience placement feature allows users to position virtual human models to simulate a real audience. These models, sourced from the Unity Asset Store for accessibility and quality, come in male and female variants. Users can create, reposition, resize, rotate, and delete these models as needed, providing flexibility in setting up a practice environment. These controls are accessible via the bottom-left panel in the application. We implemented this feature using Unity’s AR Plane Manager, Anchor Manager, and Raycast Manager to enable stable placement and interaction within the augmented space.

\par The visual aid access feature enables users to display and navigate PowerPoint slides directly within the AR environment. Users can move forward and backward through their presentation using simple controls. In this initial prototype, each slide is represented as a separate GameObject, with transitions handled manually. While this implementation is currently hardcoded, future iterations will incorporate dynamic file loading for greater flexibility.

\section{Evaluation}

We conducted a user study using a combination of Likert-scale and open-ended questions to assess participants’ past experience with SpeakAR, their confidence in public speaking, their impressions of SpeakAR, and their recommendations for improvement. Due to time constraints, we limited the study to five participants. While this sample size does not allow for statistical inference, we carefully analyzed their qualitative responses to extract meaningful insights.

\subsection{Protocol}

\par Before testing the application, we asked participants about their prior experience with XR and public speaking. Most participants had previously used XR applications, making them familiar with the technology. When asked about features that could help improve public speaking confidence, the majority emphasized the importance of an immersive experience and a realistic environment.

\par Participants completed six tasks covering key functionalities of SpeakAR, including audience placement, customization, and presentation. Table~\ref{tab:task_table} summarizes the average completion times for each task.

\par After completing the tasks, participants provided feedback on SpeakAR’s features and their overall user experience. Table~\ref{tab:task_ratings} summarizes their ratings for each of the features of the SpeakAR. Additionally, we solicited their feedback on the application. We asked questions focused on the model placement and presentation functionalities, as well as the application's perceived effectiveness in boosting confidence in public speaking. Participants rated their experience and provided suggestions for improvement.

\subsection{Findings}

\begin{table}[t]
    \captionsetup{aboveskip=3pt}
    \centering
    \caption{Task List with Average Times}
    \begin{tabular}{clr}
        \toprule
        \# & \textbf{task} & \textbf{avg time (sec)} \\ 
        \midrule
        \textbf{1} & Place down a human & 15.2 \\ 
        \textbf{2} & Enlarge the human & 6.5 \\ 
        \textbf{3} & Rotate the human & 20.1 \\ 
        \textbf{4} & Open the PowerPoint & 20.3 \\ 
        \textbf{5} & \parbox{5cm}{\vspace{0.5mm}Make the user focus on the space\\ and recite something} & 13.6 \\ 
        \textbf{6} & \vspace{0.5mm}Delete all humans & 26.6 \\ 
        \bottomrule
    \end{tabular}
    \label{tab:task_table}
\end{table}

\begin{table}[t]
 \captionsetup{aboveskip=3pt}
    \centering
    \caption{Average Rating per Question}
    \begin{tabular}{lr}
        \toprule
        \textbf{Question} & \textbf{avg rating} \\ 
        \midrule
        Effectiveness of the App & 3.4 \\ 
        Model Placing Feature & 3.2 \\ 
        Presentation Feature & 3.6 \\ 
        Necessity of the Presentation Feature & 4.6 \\ 
        Experience in Spawning Models & 3.8 \\ 
        Experience in Navigating the Presentation & 4.4 \\ 
        \bottomrule
    \end{tabular}
    \label{tab:task_ratings}
\end{table}

\par Observations during the study revealed several usability challenges. Task \#3 took longer to complete because some participants did not notice the model selection indicator and attempted to rotate an unselected model. Similarly, for Task \#4, where our prototype simulates presentation file uploads, some users did not realize the slides were already “loaded” and ready to be viewed. In Task \#6, the variation in completion times for deleting models was due to differences in the number of models each participant had spawned. Despite these challenges, participants were able to navigate the app effectively.

\par Survey responses highlighted that the virtual audience models were engaging but somewhat distracting. Participants suggested adding varied poses, animations, and dynamic facial expressions to enhance realism. The presentation feature was considered essential for effective practice. Overall, all five participants believed SpeakAR could help train and improve public speaking confidence. One participant noted, ``…it can help you pretend that there are actual people, it helps in imagining or visualization.''

\par Based on our findings, we addressed the following research questions along with the observations from our study. \textbf{RQ1}: What are the features of the application that helped the user improve their confidence? We report that the audience placement and presentation features contributed to improving user confidence. \textbf{RQ2}: How effective are the features of the application in improving user confidence? Our participants reported that the application is effective in its current state but requires further development and refinement. Finally, \textbf{RQ3}: How effective are virtual avatars as a training tool for helping our learners become comfortable to an audience while speaking? Based on our findings, augmentations can serve as an effective training tool, allowing users to practice in a simulated public environment independently, though with certain limitations.

\section{Discussion}
\par The results of the user study provided valuable insights into how users perceive virtual avatars as a tool for getting used to an audience while practicing for a speaking task. Participants generally found \textit{SpeakAR} effective in simulating a public speaking environment, with all five agreeing that it has the potential to improve confidence. One participant noted, \textit{``...it can help you pretend that there are actual people, it helps in imagining or visualization.''} This finding aligns with prior research \cite{breuss2016speaking, fitayanti2024transforming, chaidir2024effect}, which suggests that virtual audiences can enhance self-perceived preparedness and reduce anxiety in public speaking training. 

\par The model placement feature was particularly noted for enhancing realism, though some participants found the static audience models distracting. They suggested adding different poses, animations, and dynamic facial features to make the experience feel more natural. These findings are consistent with \cite{wang2019exploring}, which emphasizes the importance of audience responsiveness in immersive training environments. Similarly, the presentation feature was highly regarded, with an average rating of 4.6 out of 5, underscoring its importance in simulating real-life speaking engagements. This supports the work of \cite{visconti2023comparing}, which found that integrating visual aids in XR-based training can improve engagement and speech delivery.

\par However, despite these promising findings, a key limitation of this study is the small sample size of only five participants, which restricts the ability to generalize results. More extensive testing with a diverse set of users is necessary to fully validate these findings. Nevertheless, the initial positive responses indicate that \textit{SpeakAR} is a step in the right direction, providing a foundation for further development and refinement in XR-based public speaking training tools. Future iterations should explore the integration of real-time feedback mechanisms, as suggested by \cite{genay2021being}, to further enhance user engagement and training effectiveness.

\section{Limitations and Future Work}

\par As \textit{SpeakAR} is still in its first prototype, several refinements are needed to enhance its immersion and effectiveness in helping users improve their public speaking confidence. One major limitation of this study is the small participant pool ($n=5$), all of whom were students from a single course. Additionally, the evaluation lacked quantitative measures such as pre/post anxiety scores, making it difficult to assess the measurable impact of the system on glossophobia. Testing was also limited to a single session, which prevents conclusions about long-term effects or sustained improvements in user confidence. Broader and more structured evaluation with diverse participants and longitudinal study designs will be necessary to better understand \textit{SpeakAR}'s impact.

\par From a technical perspective, the current system lacks real-time feedback mechanisms such as speech analysis, gaze tracking, or audience response modeling—features commonly seen in more immersive VR-based systems. These functionalities are important for supporting more interactive and responsive training experiences and will be critical in future iterations of the system. In its current form, the system also does not delve deeply into AR interaction mechanics or system architecture, which may limit its extensibility and comparative value in XR research.

\par For future versions, several enhancements have been identified through user feedback. Participants emphasized the need for more realistic audience models, including varied poses, animations, and facial expressions to more closely mirror real-world public speaking scenarios. Additionally, improving the presentation feature by allowing direct integration of PowerPoint onto vertical surfaces in the AR space will contribute to a more seamless and immersive experience.

\par Further UX refinements—such as optimizing interface flow, simplifying navigation, and improving file import options—will help reduce cognitive load and improve user experience. Finally, integrating speech and audio analysis capabilities will allow the system to deliver real-time, actionable feedback on elements like pacing, volume, and clarity, making the tool more effective for skill development. By addressing these limitations and opportunities, \textit{SpeakAR} has the potential to evolve into a comprehensive and impactful AR training tool for public speaking.

\section{Conclusion}

\par We presented \textit{SpeakAR}, an augmented reality training tool for public speaking that enables users to practice in an immersive environment with virtual audience placement and integrated presentation features. Our pilot study provided initial insights into its usability and effectiveness, highlighting its potential to support confidence-building while also revealing areas for refinement. Participants found the simulated audience beneficial for visualization but suggested improvements such as more dynamic audience behaviors and enhanced UI interactions. 

\par While our findings align with prior research on XR-based public speaking training, they also emphasize the need for further development, particularly in integrating real-time speech feedback and refining presentation mechanics. Future work will focus on enhancing the realism of virtual audiences, expanding user studies to a broader demographic, and developing a more structured evaluation protocol to assess long-term effects. By addressing these challenges, \textit{SpeakAR} can evolve into a more comprehensive tool that bridges the gap between traditional rehearsal methods and immersive digital experiences.

\balance 
\bibliographystyle{ACM-Reference-Format}
\bibliography{main}

\end{document}